\documentclass[prl,twocolumn,showpacs,amsmath,amssymb,citeautoscript]{revtex4}
\usepackage{graphicx}
\usepackage{dcolumn}
\usepackage{bm}

\begin{document}

\title{Quantum Phase Transitions of Hard-Core Bosons in Background Potentials}
\author{Anand Priyadarshee} 
\author{Shailesh Chandrasekharan}
\author{Ji-Woo Lee} 
\altaffiliation{Current address:
School of Physics, Korean Institute for Advanced Study, Dongdaemun-gu, Seoul 130-722, Korea.}
\author{Harold U. Baranger}
\affiliation{ 
Department of Physics, Box 90305, Duke University,
Durham, North Carolina 27708-0305}

\date{September 2, 2006; \textbf{Phys. Rev. Lett. 97, 115703 (2006)}}

\begin{abstract}
We study the zero temperature phase diagram of hard core bosons in two dimensions subjected to three types of background potentials: staggered, uniform, and random. In all three cases there is a quantum phase transition from a superfluid (at small potential) to a normal phase (at large potential), but with different universality classes. As expected, the staggered case belongs to the $XY$ universality, while the uniform potential induces a mean field transition. The disorder driven transition is clearly different from both; in particular, we find $z \sim 1.4$, $\nu \sim 1$, and $\beta \sim 0.6$. 
\end{abstract} 

\maketitle

The ground state of a quantum system can change abruptly at certain points when a parameter in the Hamiltonian is varied. Such ``quantum phase transitions'' (QPT) have attracted a great deal of attention recently \cite{Sach}; those that involve the interplay of disorder and interactions are among the most interesting and challenging.

Superfluids and superconductors, for example, undergo a QPT to a normal state when the disorder in the medium is increased.  Experimentally such super\-fluid--normal (super\-conductor--insulator) transitions occur in $^4$He films \cite{Chan03}, arrays of Josephson junctions \cite{TakehideJJ06}, and thin-film superconductors \cite{Goldman06}. The transition is typically induced by tuning the disorder strength, chemical potential, or magnetic field. Theoretically, it is generally accepted that this transition can be studied using the bosonic Hubbard model; techniques that have been used include mean field theory \cite{MaHalpLee86,Ghosal98}, scaling and renormalization group arguments \cite{FisherWGF89,Weichman96,Herbut98,ChamonNayak02,Weichmanunpub}, and Monte Carlo calculations \cite{WallinSGY94,Zhang95,KiskerRieger97,LeeCha01,ProkSvis04,LeeCha04,AletSor03}. The transition is thought to be from the superfluid to a disordered ``Bose glass" phase.

Despite this extensive work, fundamental features of superconductor-insulator (SI) transitions due to disorder, such as their university class (or classes), remain controversial. Of particular interest in a QPT is the dynamical exponent $z$ which relates the diverging correlation lengths in temporal and spatial directions: $\xi_t \sim \xi^z$. From scaling properties, under reasonable assumptions, it was argued that $z=d$, where $d$ is the spatial dimension \cite{FisherWGF89}. Those arguments, however, have been questioned recently \cite{ChamonNayak02,Weichmanunpub}. Early studies of the bosonic Hubbard model found conflicting results for the values of $z$ and $\nu$ \cite{WallinSGY94,Zhang95}. More recently the ``current-loop'' version of this model has gained popularity \cite{KiskerRieger97,LeeCha01,ProkSvis04,LeeCha04,AletSor03}, especially for investigating the effect of weak-disorder on the clean transition near integer filling \cite{KiskerRieger97,LeeCha01,ProkSvis04,LeeCha04}, and yields results generally consistent with those from scaling theory.

In this paper we revisit the subject by providing a careful analysis of finite size scaling in and around the critical region. In particular, compared to previous work, we study much bigger systems in the superfluid phase and make no assumption concerning $z$. The new method of analysis is first applied to two clean QPT in order to demonstrate its validity. We find that the strong-disorder SI transition that we study (hard core bosons at half filling) has $z \approx 1.4$. This is strong evidence that disorder-induced SI transitions outside of the $z = d$ universality class do exist, supporting \cite{Weichmanunpub}.

\textit{Hard-core bosons in a background potential---}
We focus on a model of hard-core bosons in three different background potentials. This is the boson Hubbard model in the limit of infinite repulsion and is equivalent to a current-loop model with restricted values of the current.  The Hamiltonian is
\begin{equation}
H = - \frac{1}{2}\sum_{<ij>} [a^\dagger_i a_j + a^\dagger_j a_i] + 
\sum_i \mu_i \Big(n_i-\frac{1}{2}\Big)
\end{equation}
where $a^\dagger_i$ and $a_i$ are boson creation and annihilation operators for the site $i\equiv(i_x,i_y)$ on a $L \times L$ bipartite square lattice and $n_i = a^\dagger_i a_i$ is the number operator.  The hard core nature of the bosons constrains the boson number at each site to be $0$ or $1$. (Note that this model is equivalent to a spin-1/2 quantum XY model in a transverse magnetic field.) We consider three types of background potentials $\mu_i$: (1)\,a \textit{uniform} potential $\mu_i = \mu$ for all $i$, (2)\,a \textit{staggered} potential given by $\mu_i \!=\! \sigma_i \mu$ where $\sigma_i \!=\! 1$ (or $-1$) when $i$ belongs to the even (odd) sublattice, and (3)\,a \textit{disordered} potential given by $\mu_i$ uniformly distributed between $-\mu$ and $\mu$. Note that in the case of a staggered potential, the system is exactly at half filling for all values of $\mu$; for the disordered potential, the system is at half filling on average but not for a given realization of the disorder (except in the thermodynamic limit). 

The above model has been studied with a uniform chemical potential \cite{BernTroyer02}, as well as with a staggered potential \cite{AizenmanLieb04,Rousseau1DSL06}. However, the transition due to a random potential remains relatively unexplored by exact calculations. The destruction of superfluidity by disorder was indeed studied in \cite{Zhang95,BernardetBT02}, but the universality of the phase transition was not addressed. A nice feature of our study is that we focus on three different universality classes in the context of a single simple model. 

We use the directed-loop Monte Carlo algorithm in the discrete-time path-integral formulation \cite{SyljuSandvik02} to compute quantities. In particular we divide the Euclidean time extent $1/T$ into $M$ slices such that $\epsilon M = 1/T$. We fix $\epsilon = 0.1$ in this work. The partition function can then be written as a sum over configurations on an $L\times L\times 4M$ lattice in the occupation basis. The extra four time slices arise when only nearest neighbor interactions are allowed between two consecutive time slices.

\begin{figure}[t]
\begin{center}
\includegraphics[width=2.6in,clip]{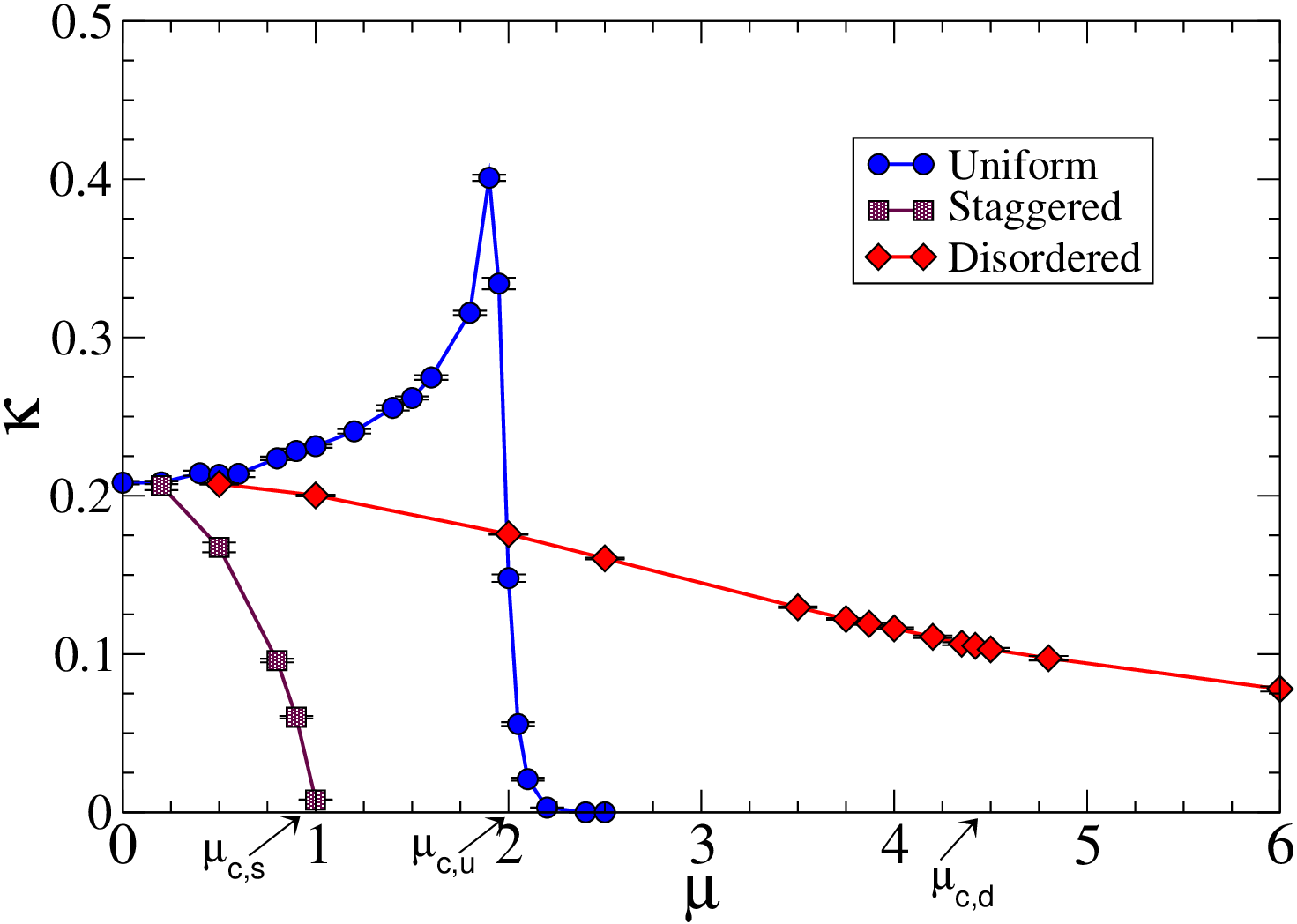}
\end{center}
\caption{\label{fig:kappa}
The compressibility as a function of the magnitude of the three different background potentials. The uniform (circles) and staggered (squares) cases are very different from the average behavior in the disordered system (diamonds).
}
\end{figure}

The different behavior caused by the three background potentials is strikingly illustrated by looking at the \textit{compressibility} as the magnitude of the potential is increased (see Fig.\,\ref{fig:kappa}). The compressibility is evaluated from
\begin{equation}
\kappa = \frac{1}{L^2T} \Big[ \Big\langle \Big( \sum_i n_i \Big)^2 \Big\rangle
                   -  \Big\langle \sum_i n_i \Big\rangle^2 \Big] \;.
\end{equation}
The SI transition is very clear in the case of the ordered potentials: it occurs at $\mu = 2$ for the uniform field and near $\mu \approx 1$ in the staggered case. At these points $\kappa$ becomes $0$ as the system develops a gap to excitations. In contrast, in the disordered case, no gap appears near the transition point, about $4.4$, and so the compressibility remains non-zero. A compressible insulating phase is the signature of a Bose glass.

\textit{Observables and scaling---} We explore the physics of the model using three observables: (1)\,The \textit{winding number susceptibility} is
\begin{equation}
\chi_w = \frac{\pi}{2} \big\langle W_x^2 + W_y^2 \big\rangle
\end{equation}
where $W_x$ and $W_y$ are the number of bosons crossing a surface perpendicular to the $x$ and $y$ directions respectively. In the thermodynamic limit this quantity is related to the superfluid density by $\rho_s = T\chi_w$.  (2)\,The \textit{order-parameter susceptibility} is
\begin{equation}
\chi_p = \frac{T}{16 L^2 M}\sum_{i,j,\tau,\tau'} C(i,\tau;j,\tau')
\end{equation}
where
$C(i,\tau;j,\tau') = \langle a_i(\tau) a^\dagger_j(\tau') \rangle$
is the boson correlation function between two space-time points, $(i,\tau)$ and $(j,\tau')$. The superfluid condensate squared is defined by $\Sigma^2 = T \chi_p/L^2$ in the
thermodynamic limit. (3)\,The \textit{spatial correlation length} $\xi_L$ is defined using the relation
\begin{equation}
G(i_x,j_x) = \sum_{i_y,\tau, \atop j_y,\tau'} \!C(i,\tau;j,\tau') 
\propto \cosh\big(\frac{|L/2-(i_x-j_x)|}{\xi_L}\big)
\end{equation}
for large spatial box size $L$ and separations $|i_x-j_x|$. We refer to the correlation length at $L=\infty$ as simply $\xi$ and compute it through a large $L$ extrapolation of $\xi_L$.

In order to access $T=0$ properties, we study the behavior of $\chi_w$, $\chi_p$, and $\xi$ as a function of $L$, $T$, and $\mu$. In the disordered case, we average over at least $50$ disorder realizations. (For smaller lattices, we use several $100$ realizations.) 
To check that this number is sufficient, for a few values of $\mu$ close to the transition we used up to $5000$ realizations. In all these cases we found that the only role of increasing the number of disorder realizations in computing the average is to reduce the error bars, which shrink as expected, i.e., as $1/\sqrt{N}$ for $N > 40$ realizations.

\begin{figure}[t]
\includegraphics[width=2.8in,clip]{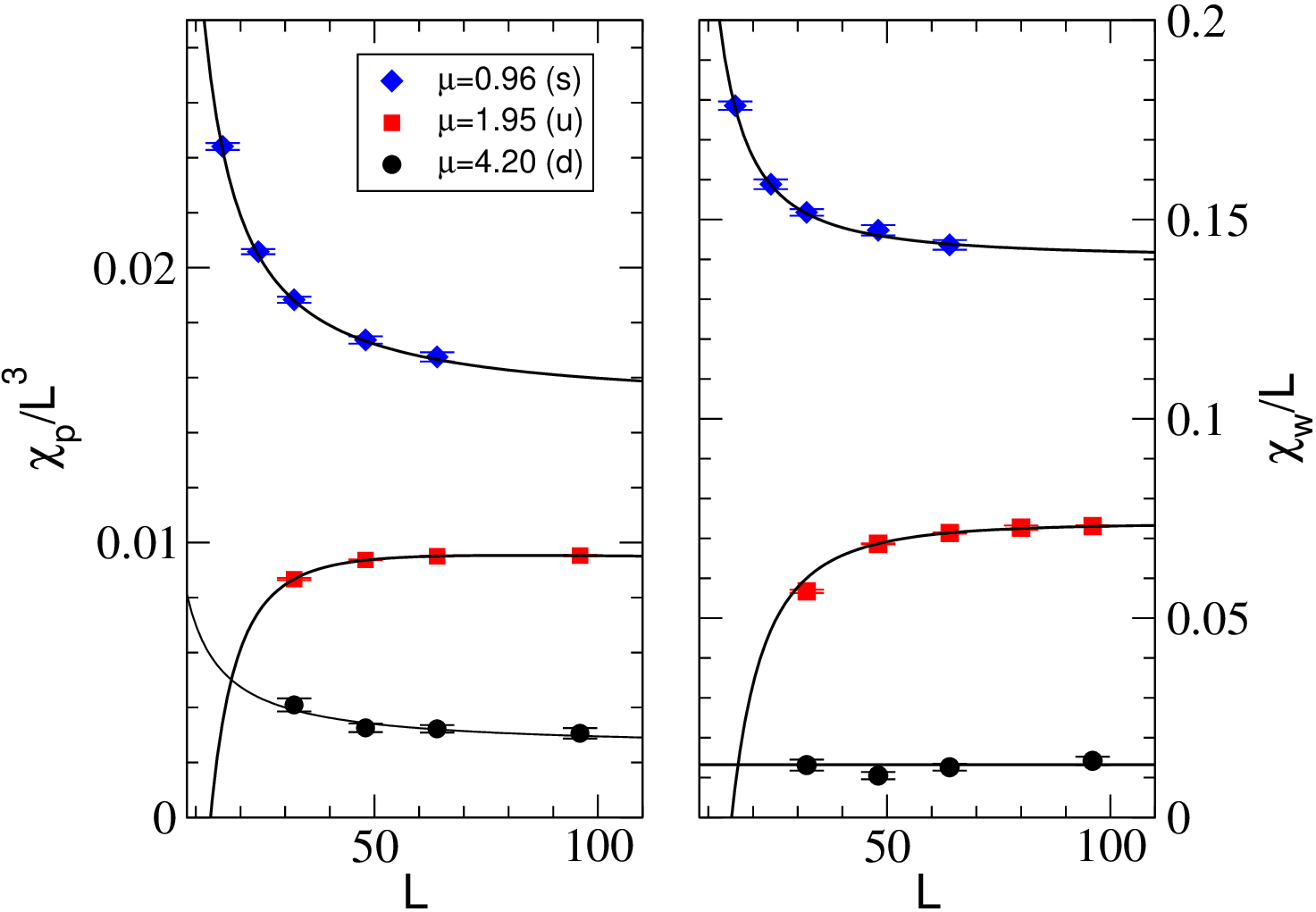}
\caption{\label{fss}
Finite size scaling of $\chi_p$ and $\chi_w$ in the superfluid phase for $\mu$ close to $\mu_c$ in a staggered, uniform, and random potential (top to bottom). The solid lines show the fits to extract $\Sigma^2$ and $\rho_s$.}
\end{figure}

For a second order phase transition, in the critical region ($\delta \equiv (\mu-\mu_c) \sim 0$, $T \rightarrow 0$, $L\rightarrow \infty$) scaling theory predicts \cite{Sach}
\begin{equation}
\chi_w = f( \delta L^{1/\nu}, T L^z), \quad
\chi_p = L^{2-\eta} g(\delta L^{1/\nu}, T L^z).
\label{scaling}
\end{equation}
When $\delta < 0$, there is a critical temperature for superfluidity such that for $T < T_c(\delta)$ and as $L\rightarrow \infty$,  $\chi_w \sim \rho_s/T$ and $\chi_p \sim \Sigma^2 L^2/T$. Consistency then constrains the scaling functions: $f(\delta L^{1/\nu},T L^z) \sim |\delta|^{z\nu}/T$ and $g(\delta L^{1/\nu}, T L^z) \sim |\delta|^{2\beta}/T$ where $2\beta = \nu(z+\eta)$. Thus, we expect $\rho_s = A |\delta|^{z\nu}$ and $\Sigma^2 = B |\delta|^{2\beta}$. In order to extract $\rho_s$ and $\Sigma^2$, we can set $1/T = L$ and study the behavior of $\chi_w$ and $\chi_p$ as a function of $L$. Note that in the superfluid phase we do not have to set $1/T \sim L^z$, making possible calculations up to $L=96$. We fit our data to the form $\chi_w = L \rho_s + a + b/L$ and $\chi_p = L^3 \Sigma^2 + c L^2 + d L$ for large $L$. Our data fits well to this expectation, allowing $\rho_s$ and $\Sigma^2$ to be  reliably extracted. Figure \ref{fss} shows this finite size scaling behavior for each of the three types of potentials at a value of $\mu$ such that $|\mu-\mu_c|/\mu_c < 5\%$.

When $\delta = 0$ scaling theory predicts that $\chi_w$ is a universal function of a single variable $T L^z$ with
$\chi_w = f_0 /(T L^z)$ for $T L^z \ll 1$ and
$\chi_w = f_1 \exp(-\alpha T^{1/z} L)$ for $T L^z \gg 1$.
We see that at $\delta=0$ the correlation length $\xi$ in the $L \rightarrow \infty$ limit is controlled by the temperature and diverges according to $\xi^{-1} = C T^{1/z}$ as $T \rightarrow 0$. We find $z$ by first extracting $\xi$ by using calculations up to $L=96$ and then fitting $\xi$ to the asymptotic form above. When $\delta > 0$, $\xi$ is expected to be finite even at $T=0$. In that regime, the relation $\xi^{-1} = D \delta^{\nu}$ defines the exponent $\nu$.

\begin{figure}[t]
\includegraphics[width=2.8in,clip]{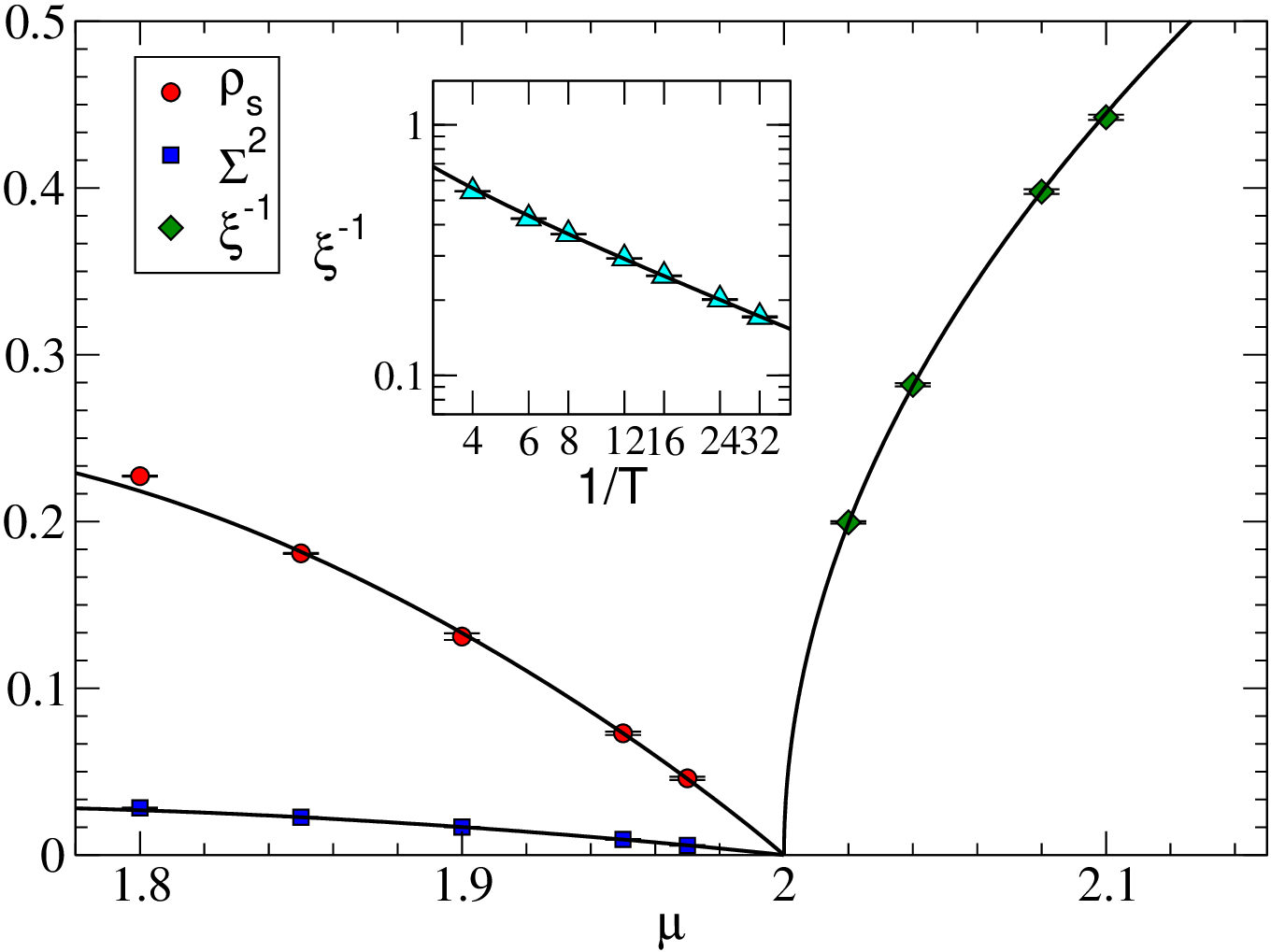}
\caption{
Uniform potential: $\rho_s$, $\Sigma^2$, and $\xi^{-1}$ 
as a function of the potential strength. The QPT is at $\mu_{c,u} = 2.0$. The inset shows the scaling of the correlation length with temperature at $\mu_{c,u}$. The solid lines are a joint fit to the expected mean field scaling.}
\label{unif}
\end{figure}

{\em Uniform background potential---} 
The QPT driven by a uniform chemical potential in a system of hard core bosons is expected to be described by mean field theory in two or more spatial dimensions \cite{FisherWGF89}. Since two spatial dimensions is the upper critical dimension, one can have large logarithmic corrections to the usual scaling forms. We assume  
$\rho_s = A |\delta|^{z\nu}(1+A'/\log\delta)$, 
$\Sigma^2 = B |\delta|^{2\beta}(1+B'/\log\delta)$,
$\xi^{-1} = C T^{1/z}(1-C'/\log T)$ and 
$\xi^{-1} = D \delta^\nu (1+ D'/\log\delta)$, 
where $\nu=1/2$, $\beta=1/2$ and $z=2$. Further, it is known \cite{BernTroyer02} that $\mu_c = 2$. Our data, shown in Fig. \ref{unif}, fits these expectations very well in the range $1.85 \leq \mu \leq 2.1$, $L > 6$ ($\chi^2/{\rm DOF} = 0.5$) \cite{uniformfit}.

\begin{figure}[t]
\includegraphics[width=2.8in,clip]{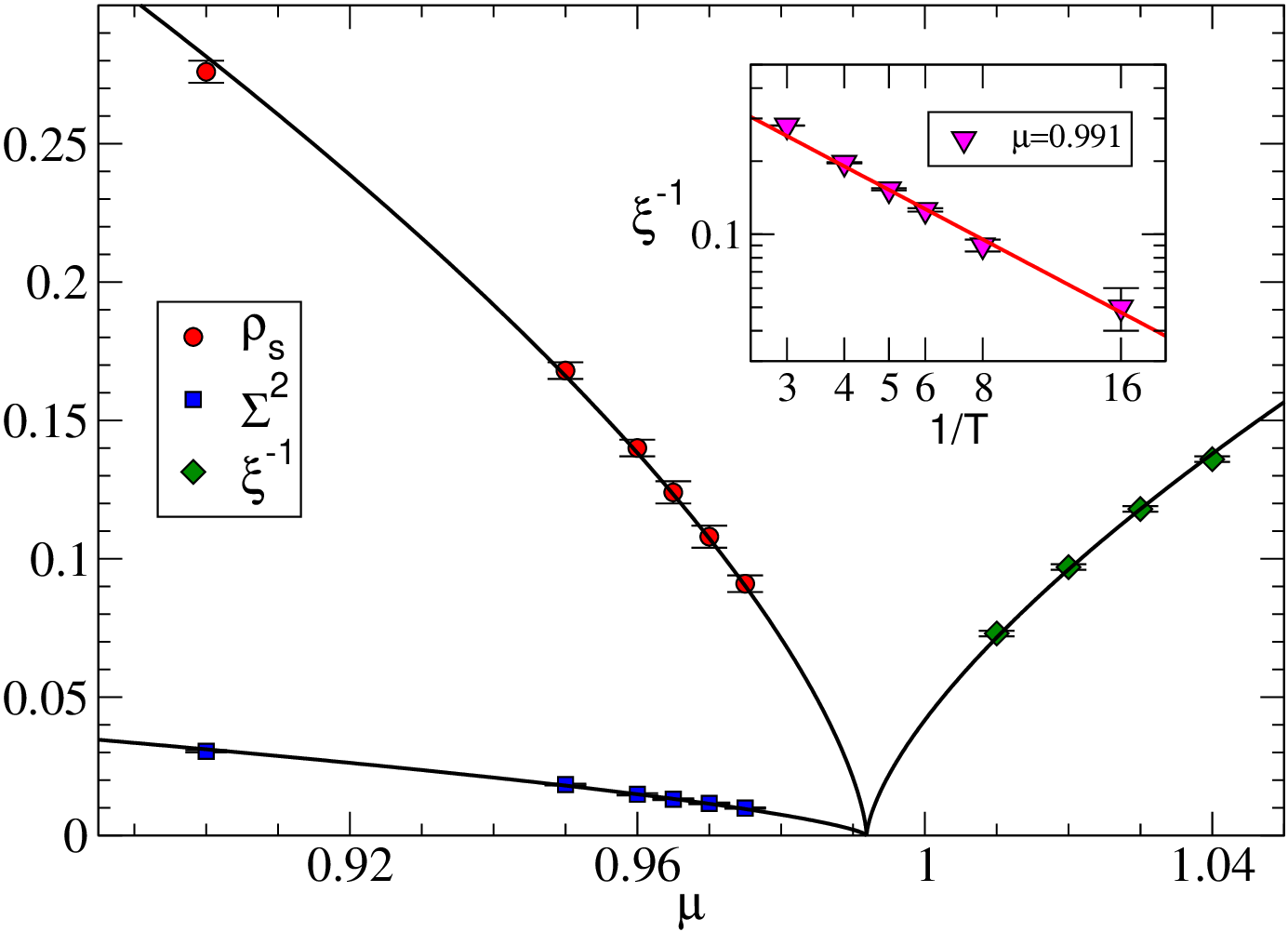}
\caption{\label{stagg}
Staggered potential: $\rho_s$, $\Sigma^2$, and $\xi^{-1}$ 
as a function of the potential strength. The QPT is at $\mu_{c,s} = 0.9919(4)$. The inset shows the scaling of the correlation length with temperature at $\mu = 0.991$. The solid lines are a joint fit to the expected 3dXY scaling.}
\end{figure}

{\em Staggered background potential---} A system of hard core bosons in a staggered chemical potential has a remaining $Z_2$ symmetry (translation by one unit followed by a particle-hole transformation). Due to this symmetry, the QPT is expected to be in the three-dimensional (3d) XY universality class with $\nu=0.672$, $\beta=0.348$, and $z=1$. With these fixed exponent values, a combined fit of all our data (except for $L \leq 4$), shown in Fig. \ref{stagg}, works extremely well \cite{stagfit}. Note that the data to demonstrate $\xi = C T^{1/z}$ (inset of Fig. \ref{stagg}) was obtained slightly off $\mu_c$; we find that this does not influence the value of $z$.

\begin{figure}[b]
\begin{center}
\includegraphics[width=2.8in,clip]{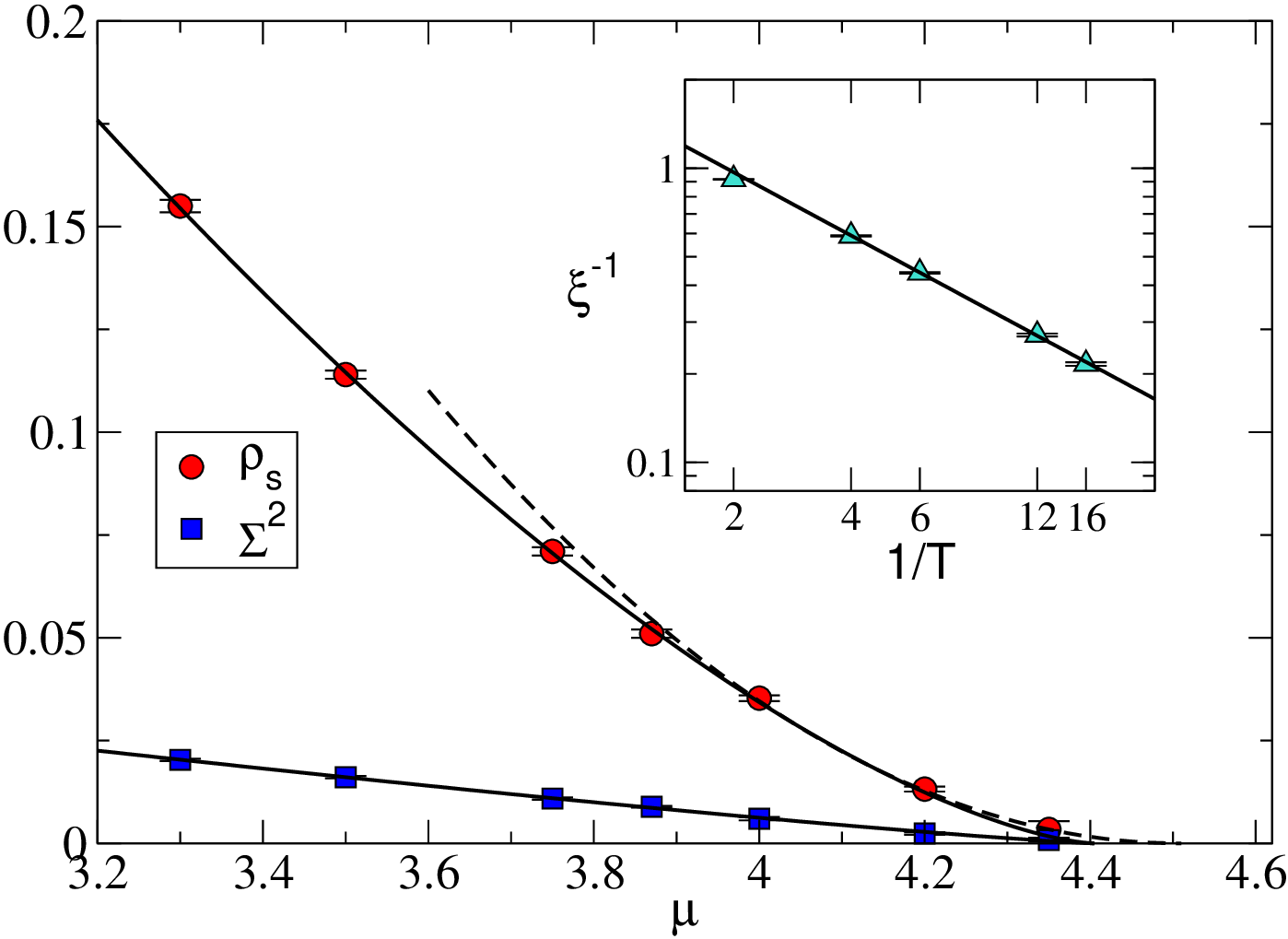}
\end{center}
\caption{\label{fig4}
Disordered potential: 
$\rho_s$ and $\Sigma^2$
as a function of the potential strength. The QPT is at $\mu_{c,d} = 4.42(2)$. The inset shows the scaling of the correlation length with temperature at $\mu=4.42$. The solid lines are a joint fit to critical behavior; the dashed line is a fit to $\rho_s = A (\mu_{c,d} - \mu)^{\nu z}$ assuming $\nu=1$ and $z=2$.}
\end{figure}

{\em Disordered potential---} Assuming that the universality class is unknown for a disordered background potential, we fit our data for $\rho_s$, $\Sigma^2$, and $\xi$, shown in Fig. \ref{fig4}, to the expected scaling forms. Note that $\xi(\mu)$ is missing: 
$\xi(T)$ appears not to saturate at small $T$ for $\mu>\mu_c$, and thus we were unable to extract $\xi(T=0)$ reliably. 

\begin{figure}[t]
\vskip0.2in
\begin{center}
\includegraphics[width=0.46\textwidth]{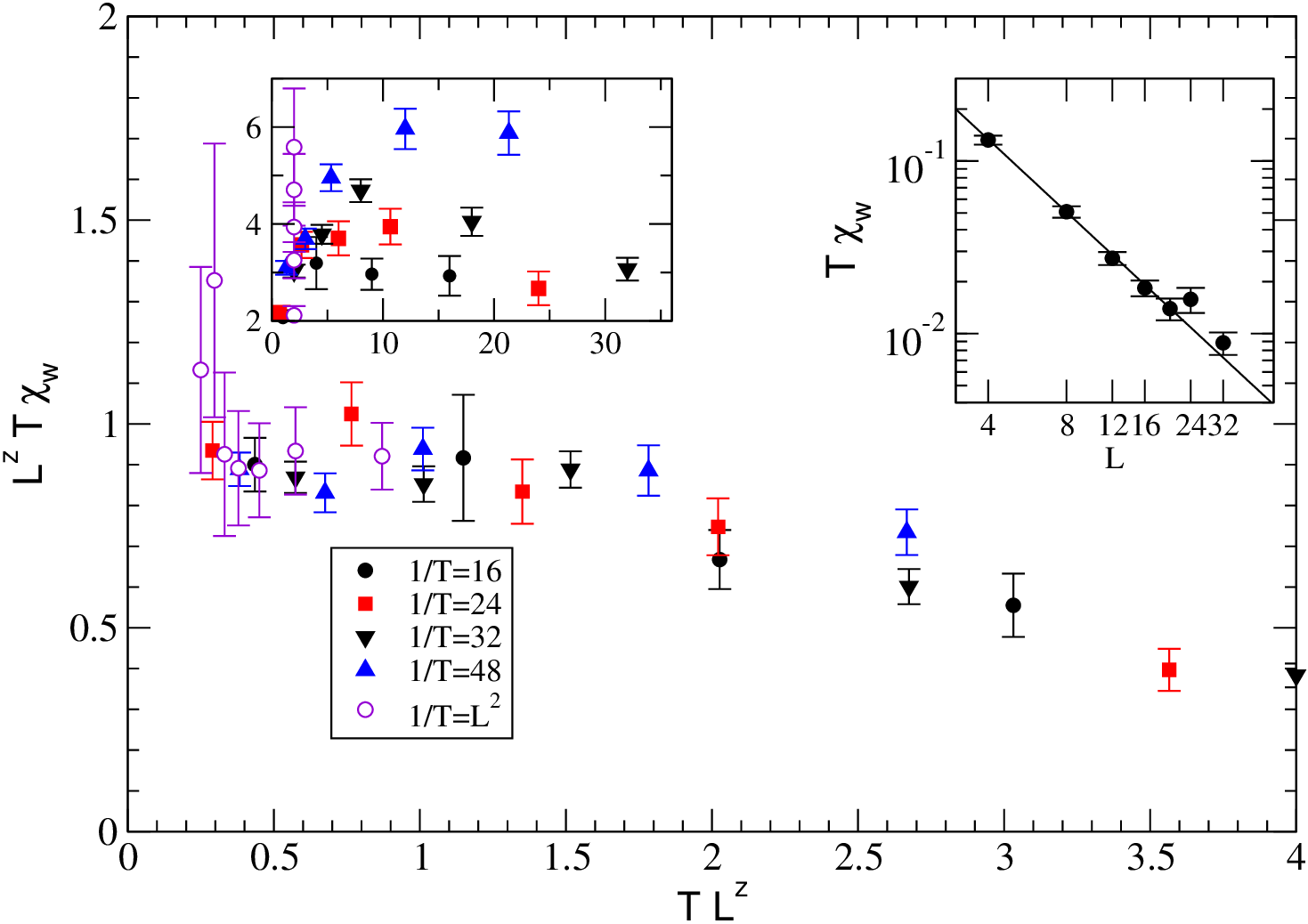}
\end{center}
\caption{\label{fig5}
Scaling function for $\chi_w$ at $\mu=\mu_c$ choosing $z=1.4$ (main figure) or $z=2$ (left inset). The right inset shows the decay of $T \chi_w$ when $T$ is scaled as $2/L^2$; the solid line is $0.93/L^{1.4}$.}
\end{figure}

Interestingly, we can fit the data for a large range, $\mu \geq 3.3$, to the scaling forms: a combined fit yields $\mu_c=4.42(2)$, $\nu = 1.10(4)$, $\beta = 0.61(2)$, and $z=1.40(2)$ with  $\chi^2/{\rm DOF} = 0.85$ (see \cite{disordfit} for prefactors). Since $2\beta = \nu(z+\eta)$, we estimate $\eta \sim -0.22(6)$. Note that $z$ is extracted by fitting $\xi^{-1} = C T^{1/z}$ at $\mu=4.42$. Although $\mu_c$ may be slightly different, we believe the value of $z$ should not be affected much, based on our experience with the staggered case. Further support for this value of $z$ is shown in Fig. \ref{fig5}: at $\mu_c$, one expects $\chi_w T L^z$ to be a function of one variable, $ T L^z$, and we see a good collapse of the data when we use $\mu = 4.42$ and $z=1.4$.

Clearly, the universality class of the phase transitions with staggered, uniform, and random potentials are strikingly different. The large size of the critical region for a disordered potential is a bit surprising. Note, however, that our results are consistent with the expectation that $\nu \geq 2/d$ \cite{Chayes86}. On the other hand, we do not see $z=2$ as predicted by \cite{FisherWGF89}. Indeed, for $z=2$, $\chi_w T L^z$ is no longer a function of $T L^z$ (inset to Fig.\,\ref{fig5}). Earlier work for strong disorder using a current loop model found $z \sim 2$ and $\nu \sim 1$ \cite{AletSor03}, implying $\rho_s = A |\delta|^2$. We can fit our data to this form (dashed line in Fig. \ref{fig4}) if we narrow the fitting region to $\mu > 4$. Such a fit yields $\mu_c = 4.49(2)$; 
This would imply that if we scale according to $T = 2/L^2$ then at $\mu = 4.42$ we must find $T\chi_w$ goes to a constant for large $L$. However, we find that $T \chi_w \sim 1/L^{1.4}$ (inset to Fig.\,\ref{fig5}) for lattices as large as $L=32$ (for which $1/T = 512$), suggesting that $\mu_c$ is close to $4.42$. In fact, even at $\mu=4.6$, $T \chi_w \sim 1/L^{1.6}$ for $T=2/L^2$ scaling. If we make the assumption that $\mu_c \leq 4.6$, then we get the bound $z \leq 1.6$.

Three possible explanations come to mind for the difference between our results and other recent results: (1)\,A fixed choice of $z$ introduced a bias in the previous analysis. In particular, a combination of small lattices and a shifted critical point may have conspired to give results consistent with $z=2$. (2)\,The disorder averages which we have performed may be insufficient: there may exist statistically significant rare events which we have not encountered. (3)\,The real critical region may be encountered only for $\mu \geq 4.35$ for which we would need lattices much larger than is possible to date.

We thank Amit Ghosal, Nandini Trivedi, and especially Peter Weichman for discussions. This work was supported in part by NSF grant DMR-0103003.

\vspace*{-0.3in}
\bibliography{dirtybosons,footnotes}

\begin{thebibliography}{27}
\bibitem[{uni()}]{uniformfit}
\bibinfo{note}{The fit to the uniform potential data is $A \!=\! 1.89(4)$, $A'
  \!=\! 0.69(3)$, $B \!=\! 0.246(5)$, $B' \!=\! 0.73(3)$, $C \!=\! 0.88(1)$,
  $C' \!=\! 0.38(3)$, $D \!=\! 1.43(1)$, and $D' \!=\! 0.04(3)$ with
  $\chi^2/{\rm DOF} \!=\! 0.5$.}
\bibitem[{sta()}]{stagfit}
\bibinfo{note}{For the staggered potential fit, we obtain $A \!=\! 1.40(2)$, $B
  \!=\! 0.164(2)$, $C \!=\! 0.762(6)$, $D \!=\! 1.06(1)$, and $\mu_c \!=\!
  0.9919(4)$ with $\chi^2/{\rm DOF} \!=\! 1.2$.}
\end{thebibliography}

\end{document}